\documentclass[%
aip,
jcp, 
 sd,%
 amsmath,amssymb,
 reprint,%
]{revtex4-1}

\usepackage{graphicx}
\usepackage{dcolumn}
\usepackage{bm}

\usepackage{tikz}
\usetikzlibrary{decorations.pathreplacing}
\usetikzlibrary{patterns}
\usetikzlibrary{calc,arrows,positioning}
\usetikzlibrary{arrows,shadows}
\usepackage{circuitikz} 

\usepackage{pgfplots}
\newcommand{\li}{\text{Li}_2\,}

\usepackage{hhline}

\newcommand{\dif}{\mathrm{d}}

\renewcommand{\i}{\mathrm{i}}

\usepackage{amsmath}	
\begin{document}

\preprint{AIP/123-QED}

\title[]
{Stroboscopic exclusion process: a first-moment-driven dynamics}
\author{Bryan Debin}
\affiliation{ 
Institut de Recherche en Math\'ematique et Physique,
Universit\'e catholique de Louvain, Louvain-la-Neuve B-1348, Belgium
} 
\author{Etienne Granet}
\affiliation{ 
The Rudolf Peierls Centre for Theoretical Physics, Oxford University, Oxford OX1 3PU, United Kingdom
} 
\date{\today}

\begin{abstract}
We define a new variant of exclusion processes in discrete time that has jump probabilities that depend on the last jump performed. In a particular limit for the jump probabilities and in suitable scaling limits for space and time, we compute the time evolution of the particle density starting from an arbitrary initial configuration, with closed boundary conditions. The core of the argument is the analysis of the time evolution of the moments. Numerical results are compared with the prediction and give excellent agreement.
\end{abstract}

\maketitle

\newcommand{\diff}[2]{\frac{d#1}{d#2}}
\newcommand{\pdiff}[2]{\frac{\partial #1}{\partial #2}}
\newcommand{\fdiff}[2]{\frac{\delta #1}{\delta #2}}
\newcommand{\bx}{\bm{x}}
\newcommand{\ba}{\bm{a}}
\newcommand{\by}{\bm{y}}
\newcommand{\bY}{\bm{Y}}
\newcommand{\bF}{\bm{F}}
\newcommand{\bn}{\bm{n}}
\newcommand{\be}{\bm{e}}
\newcommand{\new}{\nonumber\\}
\newcommand{\abs}[1]{\left|#1\right|}
\newcommand{\tr}{{\rm Tr}}
\newcommand{\HH}{{\mathcal H}}
\newcommand{\ave}[1]{\left\langle #1 \right\rangle}

\section{Introduction}

Exclusion processes are one of the most studied interacting particle systems in statistical mechanics, among which are the now classic Asymmetric Exclusion Process (ASEP)\cite{Liggett,choumallick} and its numerous variants\cite{evansferrari,prohlacevans,essler,boutillierfrancois,derridaevans,sasamoto,derridajanowsky,mallick,evansrajewsky,rajewsky,revanskafri,evansjuhasz,harrisstinchombe,harrisstinchombe2,stinchombe,janowskyl,janowskyl2,popkov,katz,vanicat}. The vast majority of the literature focuses on non-equilibrium steady states. The object of this paper is to study the relaxation towards equilibrium of yet another variant of the ASEP model. The main feature of this model is the dependence of the jump probabilities on the nature of the last jump performed. We study in particular the regime where these probabilities vary with the system size $N$ in such a way that the jumps tend to strictly alternate between left and right, in the scaling limit $N\to\infty$. The heuristics of this dynamics is that the nearly strict alternation between left and right jumps makes the evolution of the density's first moment extremely slow, so that the system is constantly on a state with maximal entropy at fixed first moment. This mechanism permits to close the hierarchy of equations usually encountered in exclusion processes. After an appropriate rescaling of the discrete time $T$,  we show that the full dynamics of the model is tractable starting from any initial state. The fact that the time evolution of the system depends on concepts of equilibrium statistical mechanics makes it particularly unusual and interesting.

\section{Definition}

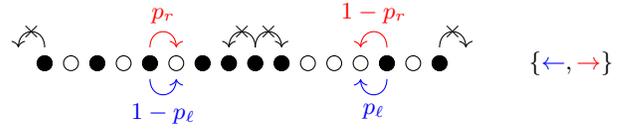
\begin{figure}
\begin{center}
\begin{tikzpicture}[scale=0.7]
\foreach \i in {0,...,15}{
\draw (\i/2,0) circle (4pt);}
\foreach \i in {0,2,4,6,7,8,9,13,15}{
\draw[fill=black] (\i/2,0) circle (4pt);}
\draw (10,0) node{$\{ {\color{blue} \leftarrow} , {\color{red} \rightarrow} \}$};
\draw [blue, ->] (2,-0.3) .. controls +(0,-0.4) and +(0,-0.4) .. (2.5,-0.3) node[midway,below]{$1-p_\ell$};
\draw [red, ->] (2,0.3) .. controls +(0,0.4) and +(0,0.4) .. (2.5,0.3) node[midway,above]{$p_r$};
\draw [blue, ->] (6.5,-0.3) .. controls +(0,-0.4) and +(0,-0.4) .. (6,-0.3) node[midway,below]{$p_\ell$};
\draw [red, ->] (6.5,0.3) .. controls +(0,0.4) and +(0,0.4) .. (6,0.3) node[midway,above]{$1-p_r$};
\draw [->] (0,0.3).. controls +(0,0.4) and +(0,0.4) .. (-0.5,0.3) node[midway]{$\times$};
\draw [->] (4,0.3).. controls +(0,0.4) and +(0,0.4) .. (3.5,0.3) node[midway]{$\times$};
\draw [->] (4,0.3).. controls +(0,0.4) and +(0,0.4) .. (4.5,0.3) node[midway]{$\times$};
\draw [->] (7.5,0.3).. controls +(0,0.4) and +(0,0.4) .. (8,0.3) node[midway]{$\times$};
\end{tikzpicture}
\vspace{-0.6cm}
\end{center}
\caption{The blue (resp. red) arrows represent the jump probabilities when the last move performed was to the left (resp. right).}
\label{ex}
\end{figure}

We consider a system of $\alpha N$ particles ($0<\alpha < 1$) located on a one-dimensional lattice of length $N$ with \textit{closed boundary conditions}. At each time step $T \to T+1$ we choose uniformly at random one of the $N$ sites. If the last move performed was to the right (resp. to the left), then with a probability $p_r$ (resp. $p_\ell$) we attempt a move to the right (resp. to the left) and with a probability $1-p_r$ (resp. $1-p_\ell$) we attempt a move to the left (resp. to the right). The attempted move is rejected: 

\begin{itemize}
\item if the departure site is not occupied by a particle, 
\item if the arrival site is already occupied,
\item if the arrival site does not exist (if we reach the extremities). 
\end{itemize}
These update rules are sketched on Figure \ref{ex}. In case of rejection, the configuration remains unchanged and a time step is still counted. By convention we fix that at $T=0$ the last particle has been moved to the right.

A few comments on this dynamics are in order. Firstly, since the dynamics depends on the previous move, it is not Markovian. However, it can easily be made Markovian if one adds an extra degree of freedom to the state of the model, taking the values in $\{\leftarrow, \rightarrow\}$ according to whether the last particle has been moved to the left or to the right, since in this formulation the dynamics depends only on the current state. Secondly, our approach can be applied to variants of this dynamics: for example if we still count a time step for an attempted but rejected move, or if the time is made continuous by introducing random ``clocks'' at each site.

On the finite lattice a configuration at a given time $T$ is entirely described by $\{X_i(T)\}_{i=1, \cdots, \alpha N}$, where $X_i(T)$ ($1 \leq X_i(T) \leq N$) is the position of the i-th particle. When we take the limit $N \to \infty$ and rescale distances by a factor $N$, the state of the system can be described (but a priori not fully) by a density $\rho(x,T)$, defined such that the number of particles at time $T$ between $N x$ and $N (x + \dif x)$ is $\rho(x,T) N \dif x $, for $0 \leq x \leq 1$. This density is equivalent to giving the discrete counting function $z_N(x,T)$, defined to be such that there are $N z_N(x,T)$ particles with position smaller than $xN$, at time $T$.

It will turn out to be fruitful to introduce the moments
\begin{equation}\label{mom}
\langle x^q \rangle (T) = \frac{1}{\alpha N} \sum_{i=1}^{\alpha N} \left( \frac{X_i(T)}{N} \right)^q\,,
\end{equation}
that can be expressed in the scaling limit $N\to\infty$ in terms of the density as
\begin{equation}
\langle x^q \rangle (T)= \frac{1}{\alpha} \int_0^1 x^q \rho(x,T) \; \dif x+o(N^0)\,.
\end{equation}

\section{Strong memory limit and separation of time scales\label{III}}

The model becomes tractable in the case where the jump probabilities $p_\ell, p_r$ depend on $N$ and get vanishingly small as $N \to \infty$. More precisely, we consider  the \textit{strong memory limit} $p_\ell, \; p_r \to 0$, as well as the \textit{weakly asymmetric regime}  $p_r-p_\ell =(p_r+p_\ell) \mathcal{O}\left(\frac{1}{N}\right)$. In the strong memory limit, it becomes increasingly likely that the sequence of jumps strictly alternates between left and right, hence the name \textit{stroboscopic exclusion process}. We will now show that in the strong memory limit, once the distances are rescaled by $N$, there are two possible rescalings of $T$ such that the rescaled system undergoes a spatial evolution of order ${\cal O}(1)$ during a ${\cal O}(1)$ time lapse. 

Let us consider two consecutive time steps in which a particle is moved and study the possible modifications of the moments \eqref{mom}. During the two time steps the quantities $\sum_i X_i^q$ for $q\geq 2$ have a probability ${\cal O}(1)$ of being modified, since the probability that the second move cancels the modification of the first one is ${\cal O}(1/N)$. However, if the two jumps are performed in opposite direction, the quantity $\sum_i X_i$ will remain the same, irrespectively of the particles selected for each jump. Hence, $\sum_i X_i$ only has a probability ${\cal O}(p_\ell+p_r)$ of being modified during these two time steps, since the second move will be in the same direction as the first one with probability $p_\ell$ or $p_r$. It follows that the $q$-th moments for $q\geq 2$ should evolve at the same rescaled time denoted $\tau$, whereas the first moment should evolve at a time scale $t=(p_\ell+p_r)\tau$. In the strong memory regime $p_\ell+p_r$ tends to zero with $N$, so that the time scale $t$ is  then infinitely slower than $\tau$. Hence, on the $\tau$ scale, the system effectively evolves at fixed first moment. These order of magnitude considerations will be confirmed in a more quantitative way in section \ref{sec_projection}.

The density of particles, i.e. the probability of observing a particle at a given position, does not fully characterize the state of the system in the scaling limit. One also requires the probability distribution of observing a certain sequence of particles and holes, which can be expressed in terms of the density only upon a mean-field approximation. In our case however, when $\tau\to\infty$ at fixed $t$ the mean field approximation is quite natural due to the scales separation and we expect it to be exact. The mean field approximation is also numerically observed to hold quite well.

To support this claim, we first recall that in the scaling limit, the evolution under time $\tau$ at fixed $t$ exactly alternates between left and right moves. Let us include the direction of the last move $\rightarrow,\leftarrow$ to the state of the system so that the evolution is Markovian. Notice that the strict alternation imposes that the reachable configurations have the same first moment, up to $\pm \frac{1}{\alpha N^2}$. With this strict alternating dynamics, the transition probability between two configurations is symmetric. It follows that the uniform probability distribution on the reachable configurations satisfies detailed balance, hence is the stationary probability distribution that is reached when $\tau\to\infty$. Entropic arguments ensure then that the configurations with a precise  density are exponentially more numerous, so that with probability $1$ in the scaling limit the system is observed in a fixed  density $\rho(x,t)$. The question is now to understand the finer details of the configurations beyond the density, namely the probability distribution of observing a certain sequence of particles and holes. We will restrict for simplicity to clusters of particles, but similar arguments can be applied to more general sequences.

To that end, we divide the interval $[0,1]$ into windows small enough so that $\rho$ can be approximated to be a constant $K/N$ on them.  
On each of these windows, since the microscopic configurations are equiprobable, the study of the sizes of the clusters of particles reduces to studying the combinatorial quantity for a fixed $q$
\begin{equation}
C_K^N=\sum_{\substack{I\subset \{1,...,N\}\\|I|=K}}h_q(I)\,,
\end{equation}
where $h_q(I)$ is the sum of $q^{m_i}$ if in the configuration $I$ the particles are grouped into clusters of size $m_1, m_2,...$. Indeed, the coefficient in front of $q^n$ in $C_K^N$ is the number of clusters of $n$ particles among all the configurations. Considering the first cluster of $b$ particles appearing after $a-1$ holes and followed by one hole, one finds the following recurrence relation
\begin{equation}
\begin{aligned}
C_K^N=\sum_{a=1}^N\sum_{b=1}^K \Big[&q^b {N-a-b\choose K-b}+C_{K-b}^{N-a-b}\\
&+\delta_{K-b,0}\delta_{N+1-a-b,0}q^b \Big]\,,
\end{aligned}
\end{equation}
from which one deduces the generating function
\begin{equation}
\sum_{K,N}C_K^N x^Ky^N=\frac{qxy(1-xy)^2}{(1-y(1+x))^2(1-qxy)}\,.
\end{equation}
From this function one finds that for $K,N\to\infty$ at $K/N$ fixed we have\cite{pemantle}
\begin{equation}
C_K^N=\frac{qK(1-K/N)^2}{1-qK/N}{N\choose K}+{\cal O}({N\choose K})\,.
\end{equation}
Upon expanding in $q$, it yields that the probability of observing a cluster of $n$ particles is $(1-\rho)^2\rho^n$, which is indeed the mean field result. Hence, the stationary state reached for $\tau\to\infty$ at fixed $t$ exactly satisfies the mean field relations.

\section{Time evolution of the moments \label{sec_time_ev}}
We now wish to determine the slowest time evolution of the system, i.e. the evolution of the system on the time scale $t$. We will see that it is entirely determined by the evolution of the first moment.

To that end, we define $\Delta_q(T)=\langle x^q \rangle (T+1) -\langle x^q \rangle (T)$ ($q \geq 1$). In what follows we will keep the $T$ dependence implicit to write more concise expressions. One has
\begin{equation}
\begin{aligned}
\Delta_q=&P_r \int_0^1 P\left(
\begin{tikzpicture}[baseline={([yshift=-.5ex]current bounding box.center)}]
\draw[fill=black] (0,0) circle (3pt) node[yshift=-0.4cm]{$xN$};
\draw (1,0) circle (3pt) node[yshift=-0.4cm]{$xN+1$}; 
\end{tikzpicture}
\right) \delta_r \langle x^q \rangle  \; \dif x \\
&+ P_\ell  \int_0^1   P\left(
\begin{tikzpicture}[baseline={([yshift=-.5ex]current bounding box.center)}]
\draw (0,0) circle (3pt) node[yshift=-0.4cm]{$xN$};
\draw[fill=black] (1,0) circle (3pt) node[yshift=-0.4cm]{$xN+1$}; 
\end{tikzpicture}
\right) \delta_\ell \langle x^q \rangle  \; \dif x\,,
\end{aligned}
\end{equation}
where the various coefficients are defined as follows. The first term corresponds to particles moved to the right and the second term to particles moved to the left. In the first term,  $P_r $ is the probability of attempting to move a particle to the right, \scalebox{0.75}{$%
P\left(
\begin{tikzpicture}[baseline={([yshift=-.5ex]current bounding box.center)}]
\draw[fill=black] (0,0) circle (3pt) node[yshift=-0.4cm]{$xN$};
\draw (1,0) circle (3pt) node[yshift=-0.4cm]{$xN+1$}; 
\end{tikzpicture}
\right)$} is the density of favourable interfaces around the rescaled position $x$, i.e. the probability that the site $xN$ is occupied with $xN+1$ free, while $\delta_r \langle x^q \rangle $ is the variation of the $q$-th moment when a particle is moved from $xN$ to $xN+1$. The factors of the second term have a similar definition.
 
We gave a compelling argument in Section \ref{III} supporting that in the strong memory limit the mean field approximation holds so that one can express the probability of interfaces as a product of observing a particle at position $xN$ times a probability of observing a hole at position $xN+1$. Hence, we have

\begin{equation}
\begin{split}
&P\left(
\begin{tikzpicture}[baseline={([yshift=-.5ex]current bounding box.center)}]
\draw[fill=black] (0,0) circle (3pt) node[yshift=-0.4cm]{$xN$};
\draw (1,0) circle (3pt) node[yshift=-0.4cm]{$xN+1$}; 
\end{tikzpicture}
\right) = \rho(x)\left(1-\rho\left(x+\frac{1}{N}\right)\right)\\
&\qquad= \rho(x)(1-\rho(x))-\frac{1}{N}\rho(x)\rho'(x) + {\cal O}(N^{-2}),\\
&P\left(
\begin{tikzpicture}[baseline={([yshift=-.5ex]current bounding box.center)}]
\draw (0,0) circle (3pt) node[yshift=-0.4cm]{$xN$};
\draw[fill=black] (1,0) circle (3pt) node[yshift=-0.4cm]{$xN+1$}; 
\end{tikzpicture}
\right) = (1-\rho(x))\rho\left(x+\frac{1}{N}\right)\\
&\qquad=\rho(x)(1-\rho(x))+\frac{1}{N}(1-\rho(x))\rho'(x) + {\cal O}(N^{-2}).
\end{split}
\end{equation}
This quantity can as well be seen as the density of interfaces and it is not necessarily uniform. One also has
\begin{equation}
\begin{split}
P_r&=p_r \langle \rightarrow \rangle + (1-p_\ell) \langle \leftarrow \rangle, \\
P_\ell&= p_\ell \langle \leftarrow \rangle + (1-p_r) \langle \rightarrow \rangle,
\end{split}
\end{equation}
where $\langle \rightarrow \rangle$ ($0 \leq \langle \rightarrow \rangle \leq 1$) is the fraction of time during which the previous move was a right jump and similarly for $\langle \leftarrow \rangle$. The factors $\delta_r \langle x^q \rangle $ and $\delta_\ell \langle x^q \rangle $ are obtained by a simple Taylor expansion, the only subtlety being that in the case of $\delta_\ell \langle x^q \rangle $ the particle is moved from $xN+1$ to $xN$. Putting everything together, we get
\begin{equation}
\begin{split}
&\Delta_q= \frac{\left[p_r \langle \rightarrow \rangle + (1-p_\ell) \langle \leftarrow \rangle \right] }{\alpha N^2}\int_0^1 \left[\rho(1-\rho)-\frac{1}{N}\rho\rho'\right]\\
&\qquad\qquad\times\left[q x^{q-1} + \frac{1}{2N} q(q-1) x^{q-2}\right] \dif x\\
& +\frac{\left[p_\ell \langle \leftarrow \rangle + (1-p_r) \langle \rightarrow \rangle \right]}{\alpha N^2} \int_0^1 \left[\rho(1-\rho)+\frac{1}{N}(1-\rho)\rho'\right] \\
&\qquad\qquad\times\left[ -q x^{q-1} -\frac{1}{2N} q(q-1) x^{q-2}\right] \dif x. 
\end{split}
\end{equation}
There remains to compute $\langle \rightarrow \rangle $ and $\langle \leftarrow \rangle $. In general, let us consider a process where we can choose between $A$, $B$ or nothing, respectively with probability $P_{A|x}$, $P_{B|x}$ and $P_{\bullet|x}$ with $x \in \{A,B\}$ the last value chosen. Pictorially we have:
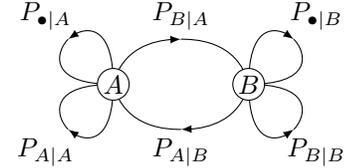
\begin{center}
\begin{tikzpicture}[>=latex, scale=0.6]
\draw [->](0,0)  .. controls +(0,-1.5) and +(0.25,-0.25) .. (-1,-1)node[below,xshift=-0.3cm]{$P_{A|A}$};
\draw [->] (-1,-1).. controls +(-0.25,0.25)  and +(-1.5,0).. (0,0) ;

\draw [->](0,0)  .. controls +(0,1.5) and +(0.25,0.25) .. (-1,1)node[xshift=-0.3cm,above]{$P_{\bullet|A}$};
\draw [->] (-1,1).. controls +(-0.25,-0.25)  and +(-1.5,0).. (0,0) ;

\draw [->] (0,0) ..controls +(0.25,1) and +(-0.25,0) .. (1.5,1) node[above]{$P_{B|A}$};
\draw (1.5,1)..controls +(0.25,0) and +(-0.25,1) .. (3,0);

\draw  (0,0) ..controls +(0.25,-1) and +(-0.25,0) .. (1.5,-1)node[below]{$P_{A|B}$};
\draw [<-](1.5,-1)..controls +(0.25,0) and +(-0.25,-1) .. (3,0);

\draw [->](3,0)  .. controls +(0,-1.5) and +(-0.25,-0.25) .. (+1+3,-1)node[below,xshift=0.3cm]{$P_{B|B}$};
\draw [->] (+1+3,-1).. controls +(0.25,0.25)  and +(1.5,0).. (3,0) ;

\draw [->](3,0)  .. controls +(0,1.5) and +(-0.25,0.25) .. (1+3,1)node[above,xshift=0.3cm]{$P_{\bullet|B}$};
\draw [->] (1+3,1).. controls +(0.25,-0.25)  and +(1.5,0).. (3,0) ;

\draw [fill=white](0,0) node{$A$} circle (10pt);
\draw [fill=white](3,0) node{$B$} circle (10pt);
\end{tikzpicture}
\end{center}
We call $\langle A \rangle $ (resp. $\langle B \rangle $) the fraction of time during which the previous choice was $A$ (resp. $B$). Then we have $\langle A \rangle + \langle B \rangle =1$ and $\langle A \rangle P_{B|A} = \langle B \rangle P_{A|B}$ so that 
\begin{equation}
\begin{split}
\langle A \rangle &= \frac{P_{A|B}}{P_{A|B} + P_{B|A}},\\
\langle B \rangle &= \frac{P_{B|A}}{P_{A|B} + P_{B|A}}.
\end{split}
\end{equation}
In the case at hand, this leads to 
\begin{widetext}
\begin{equation}
\begin{split}
\langle \rightarrow \rangle &= \frac{(1-p_\ell) \int_0^1 \left[\rho(1-\rho)-\frac{1}{N}\rho\rho'\right] \dif x}{(1-p_\ell) \int_0^1 \left[\rho(1-\rho)-\frac{1}{N}\rho\rho'\right] \dif x + (1-p_r) \int_0^1 \left[\rho(1-\rho)+\frac{1}{N}(1-\rho)\rho'\right]\dif x},\\
\langle \leftarrow \rangle &= \frac{(1-p_r) \int_0^1 \left[\rho(1-\rho)+\frac{1}{N}(1-\rho)\rho'\right] \dif x}{(1-p_\ell) \int_0^1 \left[\rho(1-\rho)-\frac{1}{N}\rho\rho'\right]\dif x + (1-p_r) \int_0^1 \left[\rho(1-\rho)+\frac{1}{N}(1-\rho)\rho'\right]\dif x}.
\end{split}
\end{equation}
Expanding this expression at first order in $1/N$ leads to
\begin{equation}
\begin{split}
\langle \rightarrow \rangle &= \frac{1-p_\ell}{2-p_r-p_\ell} - \frac{1}{N} \frac{(1-p_\ell)(1-p_r)}{(2-p_r-p_\ell)^2} \frac{\int_0^1 \rho' \, \dif x}{\int_0^1 \rho(1-\rho) \, \dif x},\\
\langle \leftarrow \rangle &= \frac{1-p_r}{2-p_r-p_\ell} + \frac{1}{N} \frac{(1-p_\ell)(1-p_r)}{(2-p_r-p_\ell)^2} \frac{\int_0^1 \rho' \, \dif x}{\int_0^1 \rho(1-\rho) \, \dif x},
\end{split}
\end{equation}
so that one finds at order ${\cal O}(N^{-3})$
\begin{equation}
\label{eq_Delta_n}
\begin{split}
\Delta_q &= \frac{1}{\alpha N^2} \frac{p_r - p_\ell}{2-p_r-p_\ell}\left( \int_0^1 \rho (1-\rho ) q x^{q-1} \, \dif x + \frac{1}{2 N} \int_0^1 \rho (1-\rho ) q(q-1)x^{q-2} \, \dif x \right)  \\
&+\frac{1}{\alpha N^3} \left[ \frac{p_\ell - p_r}{2-p_r-p_\ell} \int_0^1 \rho \rho'  q x^{q-1} \, \dif x - \frac{1-p_r}{2-p_r-p_\ell} \int_0^1 \rho'  q x^{q-1} \, \dif x \right. \\
&\left.+\frac{2(1-p_r)(1-p_\ell)(1-p_\ell-p_r)}{(2-p_r-p_\ell)^2} \frac{\int_0^1 \rho'  \, \dif x}{\int_0^1 \rho (1-\rho ) \, \dif x} \int_0^1 \rho (1-\rho)qx^{q-1}\,  \dif x\right].
\end{split}
\end{equation}
\end{widetext}

\section{Projection onto a maximal entropy state in the strong memory limit \label{sec_projection}}
In the strong memory limit, using \eqref{eq_Delta_n} one finds
\begin{equation}
\Delta_q(T) =
\left\{\begin{split}
&o(N^{-3}) \quad \text{if} \quad  q=1\\
&{\cal O}(N^{-3}) \quad \text{if} \quad q >1,
\end{split}
\right. 
\label{eq_Delta_n_order_of_magn}
\end{equation} 
where if $p_\ell-p_r = o\left(\frac{p_\ell+p_r}{N}\right)$,
\begin{equation}
\label {eq diff}
\Delta_1(T)=-\frac{p_\ell+p_r}{2 \alpha N^3} \int_0^1 \rho' \dif x \,,
\end{equation}
and if $p_\ell-p_r = {\cal O}\left(\frac{p_\ell+p_r}{N}\right)$,
\begin{equation}
\label {eq diff}
\Delta_1(T)=\frac{p_r-p_\ell}{2 \alpha N^2} \int_0^1 \rho(1-\rho) \, \dif x -\frac{p_\ell+p_r}{2 \alpha N^3} \int_0^1 \rho' \, \dif x\,.
\end{equation}
We expanded at first order in $p_{\ell},  p_{r}$ and used that in both cases $p_{\ell}-p_{r} $ is negligible compared to $p_{\ell}+p_{r} $. The asymmetry between the $q=1$ and $q >1$ cases in \eqref{eq_Delta_n_order_of_magn} has interesting consequences. If we rescale the time as 
\begin{equation}
\tau=\frac{T}{N^3}\,,
\end{equation}
for a fixed $\tau={\cal O}(1)$, the first moment is constant $\frac{\dif \langle x \rangle}{\dif \tau} = 0 $, while the other moments evolve with $\tau$, according to 
\begin{equation}
\begin{aligned}
\frac{\dif \langle x^q \rangle}{\dif \tau} = \frac{1}{2\alpha} \Big[&\frac{\int_0^1 \rho' \, \dif x}{\int_0^1 \rho(1-\rho) \, \dif x} \int_0^1 \rho(1-\rho) q x^{q-1} \; \dif x\\
&- \int_0^1 \rho' q x^{q-1} \; \dif x  \Big]\,.
\end{aligned}
\end{equation}
As for the first moment, we see that it evolves on the time scale
\begin{equation}\label{rescale}
t=\frac{T(p_r+p_\ell)}{N^3}\,.
\end{equation}
In the limit of $\tau\to \infty$ these higher moments reach their stationary states satisfying $\frac{\dif \langle x^q  \rangle}{\dif \tau} = 0 $ for $q>1$. These equations are automatically satisfied if $\partial _{x} \rho(x) \propto \rho(x) (1-\rho(x))$. 
This condition has the following interesting interpretation. 
If we suppose that in the  $\tau\to\infty$ at fixed $t$ limit, the microscopic configurations of the system with a fixed first moment are equiprobable, then the density of particles $\rho(x,t,\tau)$ when $\tau\to\infty$ at fixed $t$, reaches a $\tau$-stationary state $\rho(x,t)$ and this stationary state has to maximize the entropy functional
\begin{equation}
S[\rho]=-\int_0^1 \rho \log(\rho) + (1-\rho) \log(1-\rho) \, \dif x\,,
\end{equation}
at fixed value of the first moment $\int_0^1 x \rho \,\dif x$. Such a constrain translates into the differential equation
\begin{equation}
\label {prop to1}
\partial _{x}\rho(x,t) =-\lambda(t) \rho(x,t) \left(1-\rho(x,t) \right),
\end{equation}
with $\lambda(t)$ a Lagrange multiplier, which is indeed the equation obtained from the analysis of the time evolution of the moments.

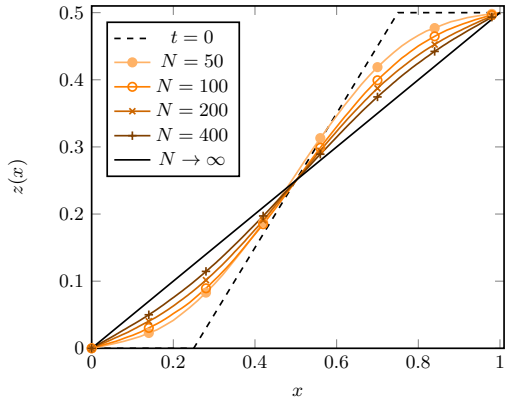
\begin{figure}[h!]
\centering
\begin{tikzpicture}[scale=0.8]
\begin{axis}[
xlabel={$x$}
,ylabel={$z(x)$}
,thick
,legend style={at={(0.36,0.5)},anchor=south east}
,xmin=0,xmax=1.01
,ymin=0,ymax=0.51
]
\addplot[black,dashed] table {relax_n200_alpha05_initial_symmetric_aver500_Nframes5_t0times0035over5.txt};
\addlegendentry{$t=0$}
\addplot[orange!60!white,mark=*,mark repeat=7] table {relax_n50_alpha05_initial_symmetric_aver500_Nframes5_t1times0035over5.txt};
\addlegendentry{$N=50$}
\addplot[orange!100!black,mark=o,mark repeat=14] table {relax_n100_alpha05_initial_symmetric_aver500_Nframes5_t1times0035over5.txt};
\addlegendentry{$N=100$}
\addplot[orange!80!black,mark=x,mark repeat=28] table {relax_n200_alpha05_initial_symmetric_aver500_Nframes5_t1times0035over5.txt};
\addlegendentry{$N=200$}
\addplot[orange!50!black,mark=+,mark repeat=56] table {relax_n400_alpha05_initial_symmetric_aver500_Nframes5_t1times0035over5.txt};
\addlegendentry{$N=400$}
\addplot[black] (0,0)--(100,500);
\addlegendentry{$N \rightarrow \infty$}
\end{axis}
\end{tikzpicture}
\caption{Average $z_N(x,t)$ over 500 configurations of various sizes for a \textit{fixed time} $t=0.007$ choosing for initial configuration an interval of $N/2$ particles at the center. The stroboscopic jump probabilities are $p_r=p_\ell=1/\sqrt{N}$. The first moment of the initial configuration vanishes, hence the system should be (instantly) projected on the state satisfying \eqref{prop to1} with vanishing first moment (solid black line). }
\label{fig_projection}
\end{figure}

Hence, if we now use the scaling \eqref{rescale} for $t$  and choose $t={\cal O}(1)$, the system is constantly projected onto a state satisfying \eqref{prop to1}. The projection process is well observed numerically as illustrated by Figure \ref{fig_projection}.

\section{Exact time evolution of the system}

\begin{figure} 
\centering
\begin{tikzpicture}[scale=0.7]
\begin{axis}[
xlabel={$x$}
,ylabel={$z(x)$}
,xtick={0,0.1,0.2,0.3,0.4,0.5,0.6,0.7,0.8,0.9,1}
,thick
,legend style={at={(0.25,0.72)},anchor=south east}
,xmin=0,xmax=1.01
,ymin=0,ymax=0.51
,every axis plot/.append style={ultra thin}
,scale=1.5
]
\addplot[black,dashed] coordinates {(0,0) (0.5,0.5) (1,0.5)};
\addlegendentry{$t=0$}
\addplot[black] table {symmetric_theor_z_at_10000t700.txt};
\addlegendentry{theory}
\addplot[only marks,mark=o,mark repeat=2,orange!50!black]  table {symmetric_n100_aver4000_z_at_10000t700.txt};
\addlegendentry{$N=100$}
\addplot[only marks,mark=x,mark repeat=4,orange!80!black] table {symmetric_n200_aver4000_z_at_10000t700.txt};
\addlegendentry{$N=200$}
\addplot[only marks,mark=+,mark repeat=6,orange!100!black] table {symmetric_n300_aver4000_z_at_10000t700.txt};
\addlegendentry{$N=300$}

\addplot[black] table {symmetric_theor_z_at_10000t1400.txt};
\addplot[only marks,mark=o,mark repeat=2,orange!50!black] table {symmetric_n100_aver4000_z_at_10000t1400.txt};
\addplot[only marks,mark=x,mark repeat=4,orange!80!black] table {symmetric_n200_aver4000_z_at_10000t1400.txt};
\addplot[only marks,mark=+,mark repeat=6,orange!100!black]  table {symmetric_n300_aver4000_z_at_10000t1400.txt};

\addplot[black] table {symmetric_theor_z_at_10000t2100.txt};
\addplot[only marks,mark=o,mark repeat=2,orange!50!black] table {symmetric_n100_aver4000_z_at_10000t2100.txt};
\addplot[only marks,mark=x,mark repeat=4,orange!80!black] table {symmetric_n200_aver4000_z_at_10000t2100.txt};
\addplot[only marks,mark=+,mark repeat=6,orange!100!black]  table {symmetric_n300_aver4000_z_at_10000t2100.txt};

\addplot[black] table {symmetric_theor_z_at_10000t2800.txt};
\addplot[only marks,mark=o,mark repeat=2,orange!50!black] table {symmetric_n100_aver4000_z_at_10000t2800.txt};
\addplot[only marks,mark=x,mark repeat=4,orange!80!black] table {symmetric_n200_aver4000_z_at_10000t2800.txt};
\addplot[only marks,mark=+,mark repeat=6,orange!100!black]  table {symmetric_n300_aver4000_z_at_10000t2800.txt};
\end{axis}

\node[fill=white] (zoom) at (7.55,2.5) {
    \begin{tikzpicture}[scale=0.6]
    \begin{axis}[
     ,xlabel={$x$}
      ,scale=0.7,
      ,enlargelimits
    ]
	\addplot[only marks,orange!50!black,mark=o,mark repeat=2] table {symmetric_n100_error_z_at_10000t2100.txt};
	\addplot[only marks,orange!80!black,mark=x,mark repeat=4] table {symmetric_n200_error_z_at_10000t2100.txt};
	\addplot[only marks,orange!100!black,mark=+,mark repeat=6] table {symmetric_n300_error_z_at_10000t2100.txt};
    \end{axis}
    \end{tikzpicture}
};

\end{tikzpicture}
\caption{Average $z_N(x,t)$ over 4 000 configurations of size $N=100, 200,300$ for $t=0.07,0.14,0.21,0.28$ choosing a step-like initial configuration with an interval of $N/2$ particles on the left. The stroboscopic jump probabilities are $p_r=p_\ell=1/\sqrt{N}$. The inset shows the absolute value of the error at $t=0.21$.}
\label{fig numerics symmetric fixed time}
\end{figure}
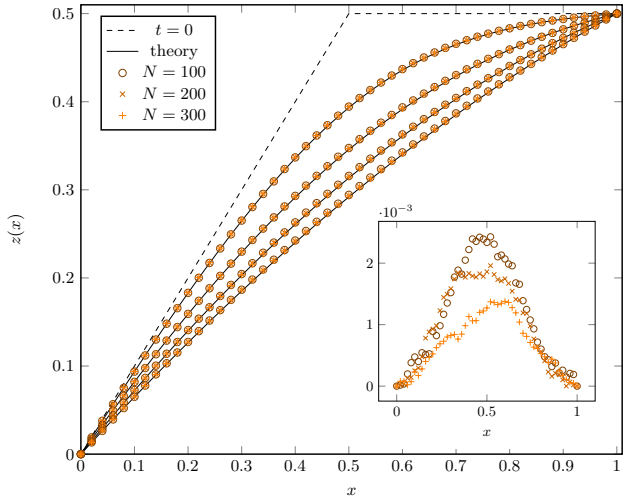

The spatial dependence is entirely determined by equation \eqref {prop to1} with the condition $\int _{0}^{1}\rho \left( x\right) dx=\alpha $. The solution is
\begin{equation}
\label{rhosol}
  \rho_\lambda (x)=\frac {1} {1+B_{\lambda }e^{\lambda x}},
\end{equation}
with
\begin{equation}
  B_{\lambda }=\frac {e^{\lambda \left( \alpha-1 \right) }-1} {1-e^{\lambda \alpha }}.
\end{equation}
Using this solution, one finds the relation between $\lambda$ and the value of the first moment to be 
\begin{equation}
\begin{aligned}
\label{eq rel lambda first moment}
  \left\langle x\right\rangle_\lambda  &=\frac {1} {2\alpha}-\frac {\left( \alpha-1\right) ^{2}} {2\alpha}+\frac {\alpha-1} {\lambda \alpha}\ln \left( \frac {1} {B_{\lambda }}+1\right) \\
  &+\frac {1} {\lambda ^{2}\alpha}\left[ \li\left( \frac {1} {1+B_{\lambda}}\right) -\li\left( \frac {1} {1+B_{\lambda}e^{\lambda}}\right) \right] .
\end{aligned}
\end{equation}
Using this relation and the differential equation for the evolution of the first moment deduced from \eqref {eq diff} gives a differential equation for $\lambda(t)$ 
\begin{equation}
\label{eqdif}
  \frac {d\lambda } {dt}=-\frac {1} {2 \alpha}\left(\frac {\beta } {\lambda }+1\right)\frac{\frac {1} {1+B_{\lambda }e^{\lambda}}-\frac {1} {1+B_{\lambda }}}{\partial_\lambda\left\langle x\right\rangle _{\lambda }},
\end{equation}
with
\begin{equation}
  \beta =\lim _{N\rightarrow \infty }N\frac {p_{r}-p_{l}} {p_{r}+p_{l}}.
\end{equation}
The initial condition for $\lambda$ depends on the initial configuration of the system described by $\rho(x,T=0)$. The initial condition $\lambda(t=0)=\lambda_0$ is a solution to 
\begin{equation}
\label{cond}
\langle x \rangle_{\lambda_0}=\frac{1}{\alpha}\int_0^1 x \rho(x,t=0) \, \dif x\,,
\end{equation}
with $\langle x \rangle_{\lambda}$ given by \eqref{eq rel lambda first moment}. Hence, in the scaling limit $N\to\infty$, starting from any initial state $\rho(x,0)$, the \textit{exact} density $\rho(x,t)$ at time $t$ is given by \eqref{rhosol}, with $\lambda(t)$ satisfying the differential equation \eqref{eqdif}, with the initial condition $\lambda(0)$ satisfying \eqref{cond}.\\

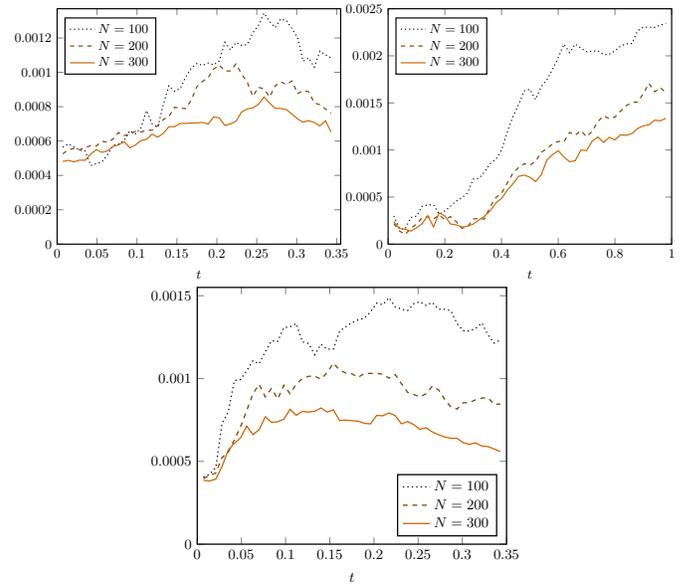
\begin{figure} 
\centering
\begin{tikzpicture}[scale=0.55]
\begin{axis}[
xlabel={$t$}
,thick
,legend style={at={(0.35,0.72)},anchor=south east}
,xmin=0,xmax=0.355
,ymin=0,ymax=0.00137
,yticklabel style={
        /pgf/number format/fixed,
        /pgf/number format/precision=7
}
,scaled y ticks=false
,xticklabel style={
        /pgf/number format/fixed,
        /pgf/number format/precision=3
}, 
]
\addplot[black,dotted] table {symmetric_n100_aver4000_fct_of_t_averabserror.txt};
\addlegendentry{$N=100$}
\addplot[orange!50!black, dashed] table {symmetric_n200_aver4000_fct_of_t_averabserror.txt};
\addlegendentry{$N=200$}
\addplot[orange!80!black] table {symmetric_n300_aver4000_fct_of_t_averabserror.txt};
\addlegendentry{$N=300$}
\end{axis}

\begin{axis}[
xshift=8cm
,xlabel={$t$}
,thick
,legend style={at={(0.35,0.72)},anchor=south east}
,xmin=0,xmax=1
,ymin=0,ymax=0.0025
,yticklabel style={
        /pgf/number format/fixed,
        /pgf/number format/precision=7
}
,scaled y ticks=false
,xticklabel style={
        /pgf/number format/fixed,
        /pgf/number format/precision=3
}, 
]
\addplot[black,dotted] table {Asymmetric_n100_aver4000_fct_of_t_averabserror.txt};
\addlegendentry{$N=100$}
\addplot[orange!50!black,dashed] table {Asymmetric_n200_aver4000_fct_of_t_averabserror.txt};
\addlegendentry{$N=200$}
\addplot[orange!80!black] table {Asymmetric_n300_aver3975_fct_of_t_averabserror.txt};
\addlegendentry{$N=300$}
\end{axis}
\end{tikzpicture}\\
\begin{tikzpicture}[scale=0.6]
\begin{axis}[
xlabel={$t$}
,thick
,legend style={at={(0.97,0.03)},anchor=south east}
,xmin=0,xmax=0.35
,ymin=0,ymax=0.00155
,yticklabel style={
        /pgf/number format/fixed,
        /pgf/number format/precision=7
}
,scaled y ticks=false
,xticklabel style={
        /pgf/number format/fixed,
        /pgf/number format/precision=3
}, 
]
\addplot[black,dotted] table {symmetric_n100_aver4000_alpha0.25_fct_of_t_averabserror.txt};
\addlegendentry{$N=100$}
\addplot[orange!50!black,dashed] table {symmetric_n200_aver4000_alpha0.25_fct_of_t_averabserror.txt};
\addlegendentry{$N=200$}
\addplot[orange!80!black] table {symmetric_n300_aver4000_alpha0.25_fct_of_t_averabserror.txt};
\addlegendentry{$N=300$}
1\end{axis}
\end{tikzpicture}
\caption{Absolute value of the error between $z_N(x,t)$ and the theoretical prediction, averaged over $x$, as a function of $t$. The functions $z_N(x,t)$ are computed by averaging over 4 000 configurations.  The stroboscopic jump probabilities are $p_r=p_\ell=1/\sqrt{N}$ (top left and bottom) and $p_r=1/\sqrt{N}+1/N^{3/2}$ and $p_\ell=1/\sqrt{N}-1/N^{3/2}$ (top right). At $t=0$ all the particles are to the left. We have $\alpha=1/2$ (top) and $\alpha=1/4$ (bottom).}
\label{fig numerics symmetric fixed time0}
\end{figure}

Let us specialize this general case to $\alpha=1/2$ and a step initial condition where all the particles are disposed to the left of the lattice. One obtains
\begin{widetext}
\begin{equation}\label{solalpha0}
\begin {aligned}
  &\rho (x, t)=\frac {1} {1+e^{\lambda (t)\left(x- \frac {1} {2}\right) }}\\
  &\frac {d\lambda} {dt}=-\frac{1}{4}\left(1+\frac{\beta}{\lambda}\right)\frac{\lambda^3\tanh (\lambda/4)}{\li\left(\frac{1}{1+e^{-\lambda/2}}\right)-\li\left(\frac{1}{1+e^{\lambda/2}}\right)+\frac{\lambda^2}{4}\frac{1}{1+e^{-\lambda/2}}-\frac{\lambda}{2}\log(1+e^{\lambda/2})}\\
  &\lambda(t=0)=+\infty\,.
  \end {aligned}
\end{equation}
\end{widetext}
This differential equation can be integrated numerically. The comparison with numerics is presented in Figures \ref{fig numerics symmetric fixed time} and \ref{fig numerics symmetric fixed time0} where we consider the quantity
\begin{equation}
z(x,t)=\int_0^x \rho(u,t) \, \dif u\,,
\end{equation}
that is the scaling limit of $z_N(x,t)$, defined before \eqref{mom}. We observe that the agreement with the numerics is excellent, with an error between $10^{-3}$ and $10^{-4}$ for the sizes considered. This error is also seen to decrease as the system size increases.

\section{Away from the strong memory limit\label{away}}
\begin{figure}
\begin{center}
\begin{tikzpicture}[scale=0.7]
\begin{axis}[
xlabel={$x$}
,ylabel={$z(x)$}
,xtick={0,0.1,0.2,0.3,0.4,0.5,0.6,0.7,0.8,0.9,1}
,thick
,legend style={at={(0.3,0.8)},anchor=south east}
,xmin=0,xmax=1.01
,ymin=0,ymax=0.51
,every axis plot/.append style={ultra thin}
,scale=1.5
]
\addplot[black,dashed] coordinates {(0,0) (0.5,0.5) (1,0.5)};
\addlegendentry{$t=0$}
\addplot[orange!60!white,mark=o,mark repeat=7] table {SSEP_n300_aver1000_z_at_10000t200.txt};
\addlegendentry{$N=300$}
\addplot[black] table {SSEP_theor_z_at_10000t200.txt};
\addlegendentry{solution \eqref{solalpha0}}
\addplot[orange!60!white,mark=o,mark repeat=7] table {SSEP_n300_aver1000_z_at_10000t400.txt};
\addplot[black] table {SSEP_theor_z_at_10000t400.txt};
\addplot[orange!60!white,mark=o,mark repeat=7] table {SSEP_n300_aver1000_z_at_10000t800.txt};
\addplot[black] table {SSEP_theor_z_at_10000t800.txt};
\addplot[orange!60!white,mark=o,mark repeat=7] table {SSEP_n300_aver1000_z_at_10000t1200.txt};
\addplot[black] table {SSEP_theor_z_at_10000t1200.txt};
\addplot[orange!60!white,mark=o,mark repeat=7] table {SSEP_n300_aver1000_z_at_10000t2000.txt};
\addplot[black] table {SSEP_theor_z_at_10000t2000.txt};
\addplot[orange!60!white,mark=o,mark repeat=7] table {SSEP_n300_aver1000_z_at_10000t4000.txt};
\addplot[black] table {SSEP_theor_z_at_10000t4000.txt};
\addplot[orange!60!white,mark=o,mark repeat=7] table {SSEP_n300_aver1000_z_at_10000t8000.txt};
\addplot[black] table {SSEP_theor_z_at_10000t8000.txt};
\end{axis}
\end{tikzpicture}
\begin{tikzpicture}[scale=0.7]
\begin{axis}[
xlabel={$x$}
,ylabel={$z(x)$}
,xtick={0,0.1,0.2,0.3,0.4,0.5,0.6,0.7,0.8,0.9,1}
,thick
,legend style={at={(0.3,0.8)},anchor=south east}
,xmin=0,xmax=1.01
,ymin=0,ymax=0.51
,every axis plot/.append style={ultra thin}
,scale=1.5
]
\addplot[black,dashed] coordinates {(0,0) (0.5,0.5) (1,0.5)};
\addlegendentry{$t=0$}
\addplot[orange!60!white,mark=o,mark repeat=7] table {WASEP_n300_aver1000_z_at_10000t200.txt};
\addlegendentry{$N=300$}
\addplot[black] table {WASEP_theor_z_at_10000t200.txt};
\addlegendentry{solution \eqref{solalpha0}}
\addplot[orange!60!white,mark=o,mark repeat=7] table {WASEP_n300_aver1000_z_at_10000t400.txt};
\addplot[black] table {WASEP_theor_z_at_10000t400.txt};
\addplot[orange!60!white,mark=o,mark repeat=7] table {WASEP_n300_aver1000_z_at_10000t800.txt};
\addplot[black] table {WASEP_theor_z_at_10000t800.txt};
\addplot[orange!60!white,mark=o,mark repeat=7] table {WASEP_n300_aver1000_z_at_10000t1200.txt};
\addplot[black] table {WASEP_theor_z_at_10000t1200.txt};
\addplot[orange!60!white,mark=o,mark repeat=7] table {WASEP_n300_aver1000_z_at_10000t2000.txt};
\addplot[black] table {WASEP_theor_z_at_10000t2000.txt};
\addplot[orange!60!white,mark=o,mark repeat=7] table {WASEP_n300_aver1000_z_at_10000t4000.txt};
\addplot[black] table {WASEP_theor_z_at_10000t4000.txt};
\addplot[orange!60!white,mark=o,mark repeat=7] table {WASEP_n300_aver1000_z_at_10000t8000.txt};
\addplot[black] table {WASEP_theor_z_at_10000t8000.txt};
\end{axis}

\end{tikzpicture}
\end{center}
\caption{Average over 1000 configurations of $z_N(xN,t)$ for $t=0.02,0.04,0.08,0.12,0.2,0.4,0.8$ and for $N=300$, starting from a step-like initial configuration with an interval of $N/2$ particles on the left. On top is shown the SSEP case with jump probabilities $p_r=p_\ell=1/2$ and on the bottom the WASEP case with $p_r=1/2+N/4$ and $p_\ell=1/2-N/4$.} 
\label{truesep}
\end{figure}
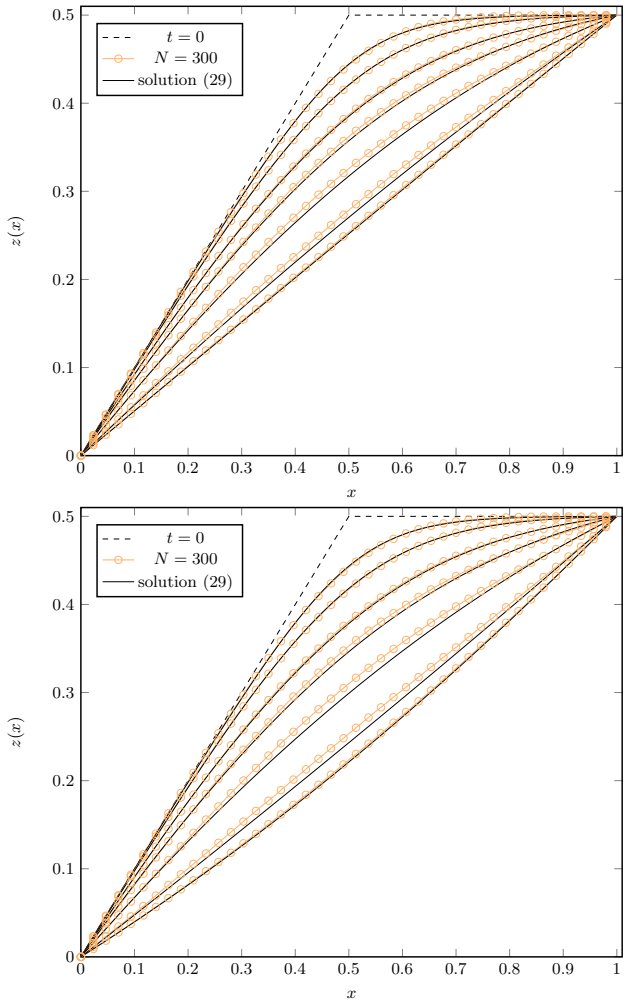

Let us now consider the more generic case where the jump probabilities are fixed to be
\begin{equation}\label{ppp}
p_r=p\left(1+\frac{a}{N}\right)\,, \qquad p_\ell=p\left(1-\frac{a}{N}\right)\,,
\end{equation}
for some values $0<p<1$ and real $a$. We note that if we set $p=1/2$, we recover the usual Weakly Asymetric Exclusion Process (WASEP) and if we further impose $a=0$ we get the Symmetric Simple Exclusion Process (SSEP). For finite $p$, a similar analysis as in Section \ref{sec_time_ev} does not exhibit a separation of time scales anymore. There is now only one macroscopic time scale, that is proportional to $N^3$. Hence, we cannot determine  the exact evolution of the model as before. However, since the strong memory limit corresponds to the limit $p\to 0$, one can consider solving \eqref{eq_Delta_n} iteratively in $p$. This would yield a series expansion
\begin{equation}\label{exprho}
\rho(x,t)=\sum_{j\geq 0}p^j \rho_j(x,t)\,,
\end{equation}
with $\rho_0(x,t)$ satisfying the differential equation \eqref{prop to1}. Then \eqref{eq_Delta_n} for $q=1$ yields
\begin{equation}\label{diff1p}
\partial_t \langle x\rangle=\frac{1}{2\alpha}\left(\frac{a}{1-p}\int_0^1\rho(1-\rho)\dif x -\int_0^1 \rho' \dif x\right)
\end{equation}
with the rescaled time
\begin{equation}
t=\frac{2Tp}{N^3}\,,
\end{equation}
as in \eqref{rescale}. We define an expansion of the full-time evolution $\rho(x,t)$ of the exclusion process with jump probabilities \eqref{ppp}, whose $n$-th term is obtained by retaining only $\rho_j$ for $j\leq n$ in \eqref{exprho}, but by keeping the full $p$ dependence in \eqref{diff1p}. Performing a similar analysis as before, we obtain that the leading term of this expansion is of the form \eqref{rhosol} with $\lambda(t)$ satisfying the differential equation
\begin{equation}
 \frac {d\lambda } {dt}=-\frac {1} {2 \alpha}\left(\frac {a } {\lambda (1-p)}+1\right)\frac{\frac {1} {1+B_{\lambda }e^{\lambda}}-\frac {1} {1+B_{\lambda }}}{\partial_\lambda\left\langle x\right\rangle}\,.
\end{equation}
In order to evaluate the precision of this first term in the expansion, we numerically compare it with the case $p=1/2$, $a=1/2$, $\alpha=1/2$, starting from a step-like initial condition, which corresponds to a standard WASEP. We also analysed the case $p=1/2$ $a=0$, $\alpha=1/2$ with a step-like initial state, which is the SSEP. The results are reported in Figure \ref{truesep} and provide a very good agreement at short times $t\lesssim 0.1$ and near the equilibrium state $t \gtrsim 0.8$ and an approximate agreement otherwise. This suggests that, if the mean field approximation still holds perturbatively in $p$, the full time evolution of the SSEP and the WASEP could be expressed as a series in $p$ from this exclusion process. Notice that in the non-strong memory limit the choice of a step-like initial configuration is crucial to make $\rho_0(x,t)$ a good approximation since the immediate projection on the states \eqref{rhosol} does not occur in this case, but instead is expected to take a time $\sim p$.

\section{Perspectives}
A natural perspective opened by this work is to study whether the approach outlined in Section \ref{away} can be pursued to describe the full time evolution of non-stroboscopic processes such as the SSEP and the WASEP. The fact that the first term of the expansion is numerically observed to approximately describe the relaxation dynamics of the  SSEP and the WASEP after a step-like initial configuration is indeed promising.

Besides, the study of this stroboscopic dynamics was inspired\cite{debin} by an attempt to describe the time evolution (according to an update rule) toward equilibrium of a directed lattice path with elementary steps $(1,0)$, $(0,1)$. Other choices of elementary steps lead to explicitly solvable Euler-Lagrange equations analogous with \eqref{prop to1}. This is the case for paths with elementary steps $(1,0)$, $(0,1)$ and $(1,1)$, for which the corresponding exclusion process consists of two types of particles. Defining an appropriate stroboscopic dynamic on this system might be interesting.

\textbf{Acknowledgments. }We thank Kirone Mallick, Hubert Saleur, Fabian Essler, Philippe Ruelle and Alexandre Lazarescu for helpful discussions. B.D. was supported by the Belgian Excellence of Science (EOS) initiative through the project 30889451 PRIMA Partners in Research on Integrable Systems and Applications. E.G. was
supported by the EPSRC under grant EP/S020527/1. Computational resources have been provided by the supercomputing facilities of the Universit\' e catholique de Louvain (CISM/UCL) and the Consortium des \' Equipements de Calcul Intensif en F\' ed\' eration Wallonie Bruxelles (C\' ECI) funded by the Fond de la Recherche Scientifique de Belgique (F.R.S.-FNRS) under convention 2.5020.11 and by the Walloon Region.

\end{document}